\documentstyle[12pt]{article}
\begin{document}

\begin{titlepage}
\begin{center} {\large \bf
MINIMIZATION OF THE SCALAR HIGGS POTENTIAL \\
IN THE FINITE SUPERSYMMETRIC GRAND UNIFIED THEORY}

\vglue 2cm {\bf I.N.~Kondrashuk}
\footnote{Supported by International
Scientific Foundation grant \#RFL000,\\
.\hspace{.5cm}E-mail: ikond@thsun1.jinr.dubna.su}

\vglue 1cm {\it Bogoliubov Laboratory of Theoretical Physics,
Joint Institute for Nuclear Research, 141 980 Dubna, Moscow Region,
RUSSIA \\} \end{center}

\vspace{2cm}

\begin{abstract}
Exact mathematical solution of the minimization conditions of scalar the
Higgs potential  of the Finite Supersymmetric Grand Unification Theory
is proposed and extremal field configurations are found. Types of
extrema are investigated and masses of the new Higgs particles arisen
after electroweak symmetry breaking are derived analytically.
The conditions for existing of physically acceptable minimum are given.
As it appears, this minimum is simple generalization of the
analogous solution in the Minimal Supersymmetric Standard Model.
Phenomenological consequences are discussed briefly.

\end{abstract}
\end{titlepage}

\section{Introduction}

Some years ago  a possibility was discovered to construct $N = 1$
supersymmetric  gauge theories with vanishing $\beta$-functions of
the gauge and Yukawa couplings in  all orders of perturbation theory
(finite theories) \cite{finite}. Following the algorithm suggested
there, a finite $SU(5)$ grand unification theory  was constructed
\cite{fgut}.  Compared to the  Minimal SUSY $SU(5)$ model, this model
has three additional pairs of Higgs multiplets. Peculiar features of the
theory are that each generation of matter interacts with its pair of
Higgs fields and each type of Yukawa interactions is degenerate with
reference to generations of fermions or, in other words, only three
different Yukawa couplings $y_U, y_D, y_L$  in accordance with three
types of interaction between Higgses and matter fields exist
\cite{fgut}.  Testing finite $SU(5)$ GUT for the compatibility with
modern precise experimental data for $sin^2 \theta_W$ and proton decay
was performed in \cite{we}. As it was shown there, these data could be
naturally reproduced within finite $SU(5)$  GUT due to  proper choosing
the mass splitting of additional multiplets in the Higgs sector of the
model.

Recently, it has been noted  that this model can possibly explain
hierarchy observed in the fermionic mass spectrum by the
hierarchy of vacuum expectation values of the Higgs
fields \cite{kaz}. To check this suggestion, it is necessary to
solve minimization conditions of the scalar Higgs potential, written on
the scale of quark masses.  From the mathematical point of view we must
solve the system of nonlinear equations, and it is a nontrivial
problem owing to a large number of Higgs fields. Nevertheless, it
appears that this system can be solved analytically. Below  we find an
exact mathematical solution of minimization conditions and analyze types
of extrema.

The paper is organized as follows. The next section is devoted to a
review of the main ideas and results of [\cite{kaz},\cite{kkk}], and
formulae, necessary for what follows, are written.  Also, we consider
the origin and special features of the scalar Higgs potential on the
$M_Z$ scale. The exact  solution of the nonlinear minimization
conditions for this potential with respect to  the neutral $SU(2)$
components of Higgs fields is offered in  section 3. The domains of
the quantities of the potential for which the extrema, presenting
interest for us from the physical point of view, exist are outlined
there too.  Further, in  chapter 4, we analyze the types of those
extremal solutions which are gauge equivalent to real field
configurations.  And, finally, we pick out the extremum which is
physically acceptable as the only candidate for a nontrivial
absolute minimum of the potential necessary for giving a fermion sector
of the theory masses.  In  conclusion, we resume peculiar features of
the potential allowing us to find  the solutions of the minimization
conditions analytically and to derive explicitly masses of the new Higgs
particles arising after the spontaneous breaking of electroweak
symmetry. The obvious phenomenological consequences for relations
between quark masses are discussed briefly.

\section{Higgs potential: origin and special features}

The multiplet contents of the unified theory has been described  in
\cite{fgut}.  For our purpose only its Higgs part is important. It
consists of four pairs of chiral superfields $\Phi_k$ and
$\overline{\Phi}_k$, $k = 1, 2, 3, 4,$ in $\bf 5$ and  $\bf \overline5$
representations of $SU(5)$, respectively, and one chiral superfield
$\Sigma$ in $24$ representation which breaks $SU(5)$ down to $SU(3)
\times SU(2) \times U(1)$. In addition to Higgs superfields, the
unified theory includes chiral matter superfields usual for the
supersymmetric $SU(5)$ theory of grand unification \cite{susysu5}.  They
are $\Psi_i$, $i = 1, 2, 3$, in $\bf \overline5$ representation and
$\Lambda_i$, $i = 1, 2, 3$, in $\bf 10$ representation of $SU(5)$, where
$i$ is generation index. The contents of these superfields have standard
form like the SUSY $SU(5)$ model \cite{susysu5}.  Reviewing the main
ideas and basic formulae of \cite{kkk} and conserving the notation used
there, we write down the Higgs and Yukawa parts of the unified finite
theory Lagrangian as

\begin{eqnarray}
{\cal L}_{Higgs + Yukawa}
&=&
y_1\Psi_iK_{ij}\overline{\Phi}_i\Lambda_j
+ y_1^{\prime}\Psi _i\overline{\Phi }_4\Lambda _i +
\frac{y_2}{8}\Phi_i\Lambda _i\Lambda _i +
\frac{y_2^{\prime}}{8}\Phi _4\Lambda _i\Lambda _i   \nonumber\\
&+&
y_3\overline{\Phi}_iS_{ij}\Sigma\Phi_j +
y_3^{\prime}\overline{\Phi }_4\Sigma \Phi _4 + \frac{y_4}{3}\Sigma ^3
\nonumber\\
&+&
\overline{\Phi }_iM_{ij}\Phi _j + \overline{\Phi }_4M\Phi _4 +
\frac{M_0}{2}\Sigma^2.
\label{1}
\end{eqnarray}
In all the terms of (\ref{1}) with repeating generation indices we
imply sum on them. The  $SU(5)$ indices are omitted here but they can
easily be restored in a covariant manner. The Yukawa constants $y_1,
y_1^{\prime}, y_2, y_2^{\prime}, y_3, y_3^{\prime}, y_4$ are expressed
in polynomial functions of the gauge coupling $g$ through the finiteness
conditions [\cite{fgut}, \cite{kkk}]. The Lagrangian is written in such
a way that only the fourth Higgs pair couples with matter like the
Higgs pair in the minimal SUSY $SU(5)$ model while other pairs
interact with each generation separately. The orthogonal matrix $S$
mixing them in the Higgs generation space will play a very important
role in this model. Its presence is not in conflict with the finiteness
conditions \cite{fgut} and a possibility to introduce it in the theory
always exists [\cite{kaz}, \cite{kkk}].  The unitary matrix $K_{ij}$ is
the usual CKM matrix \cite{fgut}.

The mass parameters of the Lagrangian (\ref{1}) are not fixed by the
finiteness conditions and by doing fine-tuning we can choose $M_0$ and
$M_{ij}$ so that $SU(5)$  should be broken in such way that we will have
only three pairs light superHiggs $SU(2)$-doublets

\begin{eqnarray}
\hat{\overline{H}}_i =
\left(
\begin{array}{c}
\hat{\overline H}_i^0 \\
\hat{\overline H}_i^{-}
\end{array}
\right) ,
\ \ \ \hat{H}_i =
\left(
\begin{array}{c}
\hat{H}_i^{+} \\
\hat{H}_i^0
\end{array}
\right), \ \ \ i = 1, 2, 3,         \label{2}
\end{eqnarray}
with opposite hypercharges (-1 and 1, respectively) below the
unification scale while the fourth pair Higgs $SU(2)$-doublets and all
colour Higgs $SU(3)$-triplets remain heavy with masses having an order
of the unification scale magnitude [\cite{kaz}, \cite{kkk}]. At this
mechanism of the $SU(5)$ violation, these three pairs of light Higgs
doublets come from those Higgs $SU(5)$-quintets that were coupled with
matter generations separately. In view of this, the Higgs part of the
Lagrangian (\ref{1}) after the $SU(5)$ symmetry violation takes the form

$$\mu_0S_{ij}\hat{\overline{H}_i}\epsilon\hat{H_j}, ~~ i,j =1,2,3, ~~
\mu_0 \sim 10^{2-3} Gev.$$
We introduce here the following notation for brevity:
$$ \overline{H}_i \epsilon H_i =
\overline{H}_i^{\alpha} \epsilon_{\alpha\beta}H_i^{\beta},$$
where $\alpha, \beta$ are the $SU(2)$ indices
\footnote{we imply that $\epsilon_{12}=1$}.

Excluding the auxiliary components of the gauge and  Higgs superfields
and adding soft supersymmetry breaking terms, we get the scalar Higgs
potential on the unification scale $M_X$ \cite{kkk}:

\begin{eqnarray}
V
&=&
(\mu_0^2 + m_0^2)|\overline{H}_i|^2 + (\mu_0^2 + m_0^2)|H_i|^2 +
(R_0\overline{H}_i \epsilon H_i +
T_0S_{ij}\overline{H}_i \epsilon H_j +h.c.)  \nonumber\\
&+&
\frac{g^2+g^{\prime}{}^2}{8}\left(|\overline{H}_i|^2-|H_i|^2\right)^2 +
\frac{g^2}{4}\left[\left(\overline{H}_i^{\dagger}\overline{H}_j\right)^{*}
\left(\overline{H}_i^{\dagger}\overline{H}_j\right) -
\left(\overline{H}_i^{\dagger}\overline{H}_i\right)^{*}
\left(\overline{H}_j^{\dagger}\overline{H}_j\right)    \right.
\nonumber \\
&+&
\left.
\left(H_i^{\dagger}H_j\right)^{*}
\left(H_i^{\dagger}H_j\right)  -
\left(H_i^{\dagger}H_i\right)^{*}
\left(H_j^{\dagger}H_j\right) +
2\left(\overline{H}_i^{\dagger}H_j\right)^{*}
\left(\overline{H}_i^{\dagger}H_j\right)\right],  \label{3}
\end{eqnarray}
where $R_0$, $T_0$ and $m_0$ are the soft breaking parameters.
Here and below we use the notation $\overline{H}_i, H_i$ for the low
scalar components of the Higgs superfields $\hat{\overline{H}_i},
\hat{H_i}.$ The gauge coupling constants $g$ and $g^{\prime}$ correspond
to the $SU(2)$ and $U(1)$ gauge group of the Standard Model,
respectively. Also we denoted for brevity $$ |\overline{H}_i|^2 =
\sum_{i} |\overline{H}_i^{\dagger}\overline{H}_i|, ~~~|H_i|^2 =
\sum_{i}|H_i^{\dagger}H_i|. $$ In other terms of (\ref{3}) we imply the
convolution of the Higgs generation indices as well. The quartic terms
in the Higgs scalar potential arise after re-expression of highest
components of the $SU(2)$ and $U(1)$ gauge supermultiplets  through
their lowest dynamical components. In this sense, the situation
is completely equivalent to that we have in the Minimal  Supersymmetric
Standard Model (MSSM)  \cite{haber}.  The difference is in
that we have three pairs of Higgs doublets instead of one in the MSSM.
It slightly complicates the form of the potential but does not result
in principal distinctions.

Below the unification scale, the finiteness property is absent and all
quantities start to renormalize while we are evolving our theory to
low energies. The remarkable property of the theory is that the quartic
terms in (\ref{3}), dictated by supersymmetry invariance, maintain their
form from high to low energies, apart from the usual renormalization of
the gauge coupling constants \cite{barbi}. The soft breaking
of supersymmetry does not alter this persistency property shown by the
exactly supersymmetric Lagrangian. On the contrary, the quadratic terms
in (\ref{3}) are slightly renormalized from their original form, and on
the $M_Z$ scale we get
\begin{eqnarray}
V &=&
m_1^2|\overline{H}_i|^2 + m_2^2|H_i|^2 + (R\overline{H}_i \epsilon H_i +
TS_{ij}\overline{H}_i \epsilon H_j +h.c.) +
\frac{g^2+g^{\prime}{}^2}{8}\left(|\overline{H}_i|^2-|H_i|^2\right)^2
\nonumber\\
&+&
\frac{g^2}{4}\left[\left(\overline{H}_i^{\dagger}\overline{H}_j\right)^{*}
\left(\overline{H}_i^{\dagger}\overline{H}_j\right) -
\left(\overline{H}_i^{\dagger}\overline{H}_i\right)^{*}
\left(\overline{H}_j^{\dagger}\overline{H}_j\right) +
\left(H_i^{\dagger}H_j\right)^{*}
\left(H_i^{\dagger}H_j\right)  \right.    \nonumber\\
&-& \left.
\left(H_i^{\dagger}H_i\right)^{*}
\left(H_j^{\dagger}H_j\right) +
2\left(\overline{H}_i^{\dagger}H_j\right)^{*}
\left(\overline{H}_i^{\dagger}H_j\right)\right]. \label{4}
\end{eqnarray}

For the spontaneous symmetry breaking to occur, this potential should
have nontrivial minimum. The vacuum expectation values of neutral
components of the $SU(2)$ doublets $H_i^0$ and $\overline{H_i^0}$ will
generate  masses  of fermions. The beauty of this model is in the
minimal influence of the Yukawa constants on the mass spectrum of the
theory.  The main load in the explanation of the observed hierarchy  in
it lies on the vacuum expectation values of the Higgs fields \cite{kaz}.
This can be shown in the following way. From the Lagrangian (\ref{1}) of
the unified theory we can get supersymmetric Yukawa Lagrangian on the
$M_X$ scale:

\begin{equation}
{\cal L}_{Yukawa} = y_DK_{ij}(Q_j \epsilon \overline{H}_i)D_i +
y_L(L_i \epsilon \overline{H}_i)E_i + y_U(Q_i \epsilon H_i)U_i,
\label{5}
\end{equation}
where $\hat{Q}$, $\hat{D}$, $\hat{U}$, $\hat{L}$, {\rm and} $\hat{E}$ are
usual matter superfields like ones in the MSSM \cite{haber}, being
$SU(2)$ doublets. If we reexpress (\ref{5}) in terms of the superfield
components, Yukawa interactions do not change their form.
One loop radiative corrections do not destroy the degeneracy
of the Yukawa constants with reference to fermionic generations
[\cite{bj}, \cite{kkk}]. Thus, on the $M_Z$ scale quarks and leptons
will gain masses
\cite{kaz}:
\begin{equation}
m_{D_i}=y_D\overline v_i,~~~
m_{U_i}=y_Uv_i,~~~m_{L_i}=y_L\overline v_i, \label{6}
\end{equation}
where  $\overline{v}_i$, $v_i$ are VEVs of  $\overline{H}^0$, $H^0$,
respectively. In view of this, it is especially important  to have the
exact solution of the minimization conditions of potential (\ref{4}).
We shall solve this problem in the next section.

\section{Solution of the minimization conditions}

For our purpose, it is convenient to rewrite our $SU(2)$ invariant
potential in  terms of the $SU(2)$ components of scalar Higgs doublets:
\begin{eqnarray}
V &=& m_1^2({\overline{H}^0_i}^{*}\overline{H}^0_i +
\overline{H}^+_i\overline{H}^-_i) +
m_2^2({H^0_i}^{*}H^0_i + H^+_iH^-_i) +
\mu_{ij}(\overline{H}^0_iH^0_j + {\overline{H}^0_i}^{*}{H^0_j}^{*})
\nonumber\\
&-&
\mu_{ij}(\overline{H}^-_iH^+_j + \overline{H}^+_iH^-_j) +
\frac{g^2+g^{\prime}{}^2}{8}\left(|\overline{H}^0_i|^2 +
\overline{H}^+_i\overline{H}^-_i -
|H^0_i|^2 - H^+_iH^-_i\right)^2
\nonumber\\
&+&
\frac{g^2}{2}\left(
\overline{H}^0_i{\overline{H}^0_j}^{*}\overline{H}^+_i\overline{H}^-_j -
\overline{H}^0_i{\overline{H}^0_i}^{*}\overline{H}^+_j\overline{H}^-_j
+ H^0_i{H^0_j}^{*}H^+_jH^-_i - H^0_i{H^0_i}^{*}H^+_jH^-_j \right.
\nonumber\\
&+&   \left.
\overline{H}^0_i{\overline{H}^0_i}^{*}H^+_jH^-_j +
\overline{H}^0_iH^0_j\overline{H}^+_iH^-_j +
{\overline{H}^0_i}^{*}{H^0_j}^{*}\overline{H}^-_iH^+_j +
{H^0_i}^{*}H^0_i\overline{H}^+_j\overline{H}^-_j\right),  \label{7}
\end{eqnarray}
where $\overline{H}^+_i = (\overline{H}^-_i)^{*}$,
$H^-_i = (H^+_i)^{*}$  and $\mu_{ij} = R\delta_{ij} + TS_{ij}$.

It is  necessary for us to find the nontrivial extremum of this
potential with reference to neutral components, and conditions which
must be satisfied  for its existence.
For this aim, we need to solve
the system of nonlinear equations
\begin{eqnarray}
\frac{1}{2}\frac{\delta V}{\delta\overline{H}_i}
&=&
m_1^2\overline{H}_i + \mu_{ij}H_j +
\frac{g^2+g^{\prime}{}^2}{4}\left(\overline{H}_i^2 +
\overline{h}_i^2 - H_i^2 - h_i^2\right)\overline{H}_i = 0 \nonumber\\
\frac{1}{2}\frac{\delta V}{\delta{H_i}}
&=&
m_2^2H_i + \mu_{ji}\overline{H}_j -
\frac{g^2+g^{\prime}{}^2}{4}\left(\overline{H}_i^2 +
\overline{h}_i^2 - H_i^2 - h_i^2\right)H_i = 0 \nonumber\\
\frac{1}{2}\frac{\delta V}{\delta\overline{h}_i}
&=&
m_1^2\overline{h}_i - \mu_{ij}h_j +
\frac{g^2+g^{\prime}{}^2}{4}\left(\overline{H}_i^2 +
\overline{h}_i^2 - H_i^2 - h_i^2\right)\overline{h}_i = 0 \nonumber\\
\frac{1}{2}\frac{\delta V}{\delta{h_i}}
&=&
m_2^2h_i - \mu_{ji}\overline{h}_j -
\frac{g^2+g^{\prime}{}^2}{4}\left(\overline{H}_i^2 +
\overline{h}_i^2 - H_i^2 - h_i^2\right)h_i = 0,     \label{8}
\end{eqnarray}
where we introduced the new notation for brevity:
$$\overline{H}_i = Re \overline{H}^0_i, ~~~
\overline{h}_i = Im \overline{H}^0_i,  ~~~
H_i = Re H^0_i, ~~~ h_i = Im H^0_i.$$
Further, we shall also denote
$${\cal H}_i = H_i + \imath h_i, ~~~ {\cal\overline{H}}_i =
\overline{H}_i + \imath \overline{h}_i.$$

 Although these equations are written in terms of the real and imaginary
parts of the neutral components of $SU(2)$ doublets, it can easily be
seen that they are invariant under the abelian gauge transformations
$${\cal H}_i \rightarrow e^{\imath\alpha}{\cal H}_i, ~~~~
{\cal\overline{H}}_i \rightarrow
e^{-\imath\alpha}{\cal\overline{H}}_i.$$
As we can see, this system
contains nonlinearity as a quadratic combination, whose square was
in the potential (\ref{7}).  It is the key property of system allowing
us to solve it analytically.  As a first step, let us rewrite
(\ref{8}) in the matrix form denoting the quadratic combination by $x$:
\begin{eqnarray}
&&(m_1^2 + x)\overline{H} + \mu H = 0 \nonumber\\
&&(m_2^2 - x)H + \mu^{T} \overline{H} = 0 \nonumber\\
&&(m_1^2 + x)\overline{h} - \mu h = 0 \nonumber\\
&&(m_2^2 - x)h - \mu^{T} \overline{h} = 0 \nonumber\\
&&x = \frac{g^2+g^{\prime}{}^2}{4}\left(\overline{H}^2 +
\overline{h}^2 - H^2 - h^2\right) \label{9},
\end{eqnarray}
where $H,~\overline{H},~h,~{\rm and}~\overline{h}$ are
the real vectors in the
Higgs generation space:
\begin{eqnarray}
\overline{H} =
\left(
\begin{array}{c}
\overline{H}_1 \\
\overline{H}_2 \\
\overline{H}_3 \\
\end{array}
\right),~~~
H =
\left(
\begin{array}{c}
H_1 \\
H_2 \\
H_3 \\
\end{array}
\right),~~~
\overline{h} =
\left(
\begin{array}{c}
\overline{h}_1 \\
\overline{h}_2 \\
\overline{h}_3 \\
\end{array}
\right),~~~
h =
\left(
\begin{array}{c}
h_1 \\
h_2 \\
h_3 \\
\end{array}
\right), \label{10}
\end{eqnarray}
and $\mu$ is matrix with elements $\mu_{ij}$ operating in the
generation space. It can be found that
$$det \mu = (R + T)(R^2 + T^2 + RT(trS -1)).$$
Below we shall suggest $R \ne 0$, $T \ne 0$, $det \mu \ne 0.$
Now let us reduce (\ref{9}) to an equivalent system
\begin{eqnarray}
&&\mu\mu^{T}H = (m_1^2 + x)(m_2^2 - x)H \nonumber\\
&&\overline{H} = - (\mu^{T})^{-1}(m_2^2 - x)H \nonumber\\
&&\mu\mu^{T}h = (m_1^2 + x)(m_2^2 - x)h \nonumber\\
&&\overline{h} = (\mu^{T})^{-1}(m_2^2 - x)h. \label{11}
\end{eqnarray}
It is  obvious that if system (\ref{11}) has a nontrivial
solution, the condition  \begin{equation}
det(\mu\mu^{T} - (m_1^2 +x)(m_2^2 - x)I) = 0 \label{12}
\end{equation}
should be satisfied. It is equivalent to
\begin{equation}
(m_1^2 + x)(m_2^2 - x) = \lambda RT + (R^2 + T^2), \label{13}
\end{equation}
where $\lambda$  is the eigenvalue of the matrix $S + S^{T}$ :
\begin{equation}
det(S + S^{T} - \lambda I) = - (\lambda - 2)(\lambda - (trS - 1))^2.
\label{14}
\end{equation}
We shall have two variants of the solution of the system, depending on
what eigenvalue is taken into account. Let us consider both the
cases.

\subsection{Solution in the $\lambda = 2$ case}

In this case, the system (\ref{11}) has the form
\begin{eqnarray}
&&(S + S^{T})H = 2H \nonumber\\
&&(S + S^{T})h = 2h \nonumber\\
&&\overline{H} = - (\mu^{T})^{-1}(m_2^2 - x)H \nonumber\\
&&\overline{h} = (\mu^{T})^{-1}(m_2^2 - x)h \nonumber\\
&&(m_1^2 + x)(m_2^2 - x) = (R + T)^2. \label{15}
\end{eqnarray}
Solutions of the first and second equations are:
\begin{eqnarray}
H = k_1\left(
\begin{array}{c}
s_{23}-s_{32} \\
s_{31}-s_{13} \\
s_{12}-s_{21} \\
\end{array}
\right), ~~~~
h = k_2\left(
\begin{array}{c}
s_{23}-s_{32} \\
s_{31}-s_{13} \\
s_{12}-s_{21} \\
\end{array}
\right). \label{16}
\end{eqnarray}
 The fifth equation of (\ref{15}) puts the restriction on  $k_1$ and
$k_2$. In fact, from (\ref{15}) we get
\begin{equation}
x = \frac{1}{2}\left(m_2^2 - m_1^2 \pm \sqrt{(m_1^2 + m_2^2)^2 - 4(R
+T)^2}\right) \label{17}.
\end{equation}
On the other side
$$x = \frac{g^2+g^{\prime}{}^2}{4}\left(\overline{H}^2 +
\overline{h}^2 - H^2 - h^2\right) =
\frac{g^2+g^{\prime}{}^2}{4}\left(H^2 + h^2\right)\left(\frac{(m_2^2 -
x)^2}{(R + T)^2} - 1\right).$$
Hence,
\begin{equation}
H^2 + h^2 = \left(m_1^2 + m_2^2 \pm  \sqrt{(m_1^2 + m_2^2)^2 - 4(R
+T)^2}\right)F_{\pm}\left((R +T)^2\right), \label{18}\\
\end{equation}
\begin{equation}
\overline{H}^2 + \overline{h}^2 = \left(m_1^2 + m_2^2 \mp  \sqrt{(m_1^2
+ m_2^2)^2 - 4(R +T)^2}\right)F_{\pm}\left((R +T)^2\right), \label{19}\\
\end{equation}
where
$$F_{\pm}(\kappa) = \frac{1}{g^2 + g^{\prime}{}^2} \frac{\pm ( m_1^2 -
m_2^2) - \sqrt{(m_1^2 + m_2^2)^2 - 4\kappa}}{\sqrt{(m_1^2 + m_2^2)^2 -
4\kappa}}.$$
 From (\ref{18}) we can easily get
$$ k_1^2 + k_2^2 =
\frac{\left(m_1^2 + m_2^2 \pm  \sqrt{(m_1^2 + m_2^2)^2 - 4(R
+T)^2}\right)F_{\pm}\left((R +T)^2\right)}{4 - (trS - 1)^2}.$$
Using the parametrization $k_1 = k\cos\phi, k_2 = k\sin\phi,$
we can write the solution of (\ref{15}) for $H$ and $h$ as
\begin{eqnarray}
H &=& \cos\phi
\sqrt{\frac{\left(m_1^2 + m_2^2 \pm  \sqrt{(m_1^2 + m_2^2)^2 - 4(R
+T)^2}\right)F_{\pm}\left((R +T)^2\right)}{4 - (trS - 1)^2}}
\left(
\begin{array}{c}
s_{23}-s_{32} \\
s_{31}-s_{13} \\
s_{12}-s_{21} \\
\end{array}
\right),  \nonumber \\
h &=& \sin\phi
\sqrt{\frac{\left(m_1^2 + m_2^2 \pm  \sqrt{(m_1^2 + m_2^2)^2 - 4(R
+T)^2}\right)F_{\pm}\left((R +T)^2\right)}{4 - (trS - 1)^2}}
\left(
\begin{array}{c}
s_{23}-s_{32} \\
s_{31}-s_{13} \\
s_{12}-s_{21} \\
\end{array}
\right). \nonumber\\
\label{20}
\end{eqnarray}

It is obvious that if
${\cal H}_i = H_i + \imath h_i$  is a solution of the system (\ref{15}),
${\cal H}_i^{\prime} = e^{\imath\alpha}{\cal H}_i$ will be its
solution as well. It is a consequence of the abelian symmetry of
equations (\ref{8}). We need to replace only the angle $\phi$ in
(\ref{20}) on the angle $\phi + \alpha$. As it can be seen, all
solutions (\ref{20}) are gauge equivalent to the real ones. However,
>from the physical point of view, we are interested in the solutions
equivalent to the real positive field configurations. It constrains the
matrix $S$, which is a parameter of the theory, because to attain it we
must choose such $S$ that all components of the vector
$\epsilon_{ijk}S_{jk}$, around which the matrix $S$ rotates all other
vectors of the three dimensional generation space, have the same signs.
In addition to these constraints there are others. In order to get
real and positive right-hand sides of (\ref{18}) and (\ref{19}) and to
have the potential bounded from below in the direction of vanishing
quartic terms in (\ref{7}), the following conditions should be
satisfied :
\begin{eqnarray}
&& m_1^2 + m_2^2 > 2|R + T|, \label{21} \\
&& m_1^2m_2^2 < (R + T)^2. \label{22}
\end{eqnarray}
Arbitrariness in  choosing signs in (\ref{20})
originating from (\ref{17}) is fixed in the following way : we take upper
sign if $m_1^2 > m_2^2$ and lower sign in the opposite case.  Knowing
(\ref{20}) we can get from (\ref{15})
\begin{eqnarray} \overline{H} &=&
- \cos\phi~sign(R +T) \sqrt{\frac{\left(m_1^2 + m_2^2 \mp  \sqrt{(m_1^2
+ m_2^2)^2 - 4(R +T)^2}\right)}{4 - (trS - 1)^2}} \times \nonumber\\
&& \times \sqrt{F_{\pm}\left((R +T)^2\right)} \left( \begin{array}{c}
s_{23}-s_{32} \\ s_{31}-s_{13} \\
s_{12}-s_{21} \\ \end{array}
\right),  \nonumber \\ \overline{h} &=&
\sin\phi~sign(R +T) \sqrt{\frac{\left(m_1^2 + m_2^2 \mp  \sqrt{(m_1^2 +
m_2^2)^2 - 4(R +T)^2}\right)}{4 - (trS - 1)^2}} \times \nonumber\\
&& \times \sqrt{F_{\pm}\left((R +T)^2\right)} \left(
\begin{array}{c} s_{23}-s_{32} \\
s_{31}-s_{13} \\ s_{12}-s_{21} \\
\end{array} \right).  \label{23}
\end{eqnarray}
It is necessary to have such $R$ and $T$ that $(R + T) <
0$. Otherwise we can not make solutions (\ref{20}) and (\ref{23}) real
and positive simultaneously.

\subsection{Solution in the $\lambda = trS - 1$ case}

In the same manner we can analyze  the second variant of the extremal
solution when $\lambda = trS - 1$.  Instead of (\ref{15}) we  get:
\begin{eqnarray}
&&(S + S^{T})H = (trS -1)H \nonumber\\
&&(S + S^{T})h = (trS - 1)h \nonumber\\
&&\overline{H} = - (\mu^{T})^{-1}(m_2^2 - x)H \nonumber\\
&&\overline{h} = (\mu^{T})^{-1}(m_2^2 - x)h \nonumber\\
&&(m_1^2 + x)(m_2^2 - x) = R^2 + T^2 + RT(trS - 1). \label{24}
\end{eqnarray}
It is easy to show that the solution of the first equation is the vector
$H$ satisfying the equation
\begin{equation}
(s_{23}-s_{32})H_1 + (s_{31}-s_{13})H_2 + (s_{12}-s_{21})H_3 = 0.
\label{25}
\end{equation}
This is true for the second equation of (\ref{24}) too. The general
solution of the first two equations is
\begin{eqnarray}
H &=& \left(
\begin{array}{c}
- K_1(s_{31}-s_{13}) - K_2(s_{12}-s_{21})\\
K_1(s_{23} - s_{32}) \\
K_2(s_{23} - s_{32}) \\
\end{array}
\right),  \label{26} \\
h &=& \left(
\begin{array}{c}
- k_1(s_{31}-s_{13}) - k_2(s_{12}-s_{21})\\
k_1(s_{23} - s_{32}) \\
k_2(s_{23} - s_{32}) \\
\end{array}
\right),  \label{27}
\end{eqnarray}
where $K_1,~K_2,~k_1,~{\rm and}~k_2$ are some quantities. The fifth
equation of (\ref{24}) puts a constraint on them.
In fact, from it we find
\begin{equation}
x = \frac{1}{2}\left(m_2^2 - m_1^2 \pm \sqrt{(m_1^2 + m_2^2)^2 - 4(R^2
+ T^2 + RT(trS - 1))}\right) \label{28}
\end{equation}
and
\begin{eqnarray}
H^2 + h^2 &=& \left(m_1^2 + m_2^2 \pm  \sqrt{(m_1^2 + m_2^2)^2 - 4(R^2
+ T^2 + RT(trS - 1))}\right) \times \nonumber\\
&& \times F_{\pm}\left(R^2 + T^2 + RT(trS - 1)\right), \label{29} \\
\overline{H}^2 + \overline{h}^2 &=& \left(m_1^2 + m_2^2 \mp
\sqrt{(m_1^2 + m_2^2)^2 - 4(R^2 + T^2 + RT(trS - 1))}\right) \times
\nonumber\\
&& \times F_{\pm}\left(R^2 + T^2 + RT(trS - 1)\right).
\label{30}
\end{eqnarray}
To get the real and positive  right-hand sides of (\ref{29}) and
(\ref{30})
and to have the potential bounded from
below in the direction of vanishing quartic terms in  (\ref{7}), the
following conditions should be satisfied :
\begin{eqnarray}
&& m_1^2 + m_2^2 > 2\sqrt{R^2 + T^2 + RT(trS - 1)}, \label{31} \\
&& m_1^2m_2^2 < R^2 + T^2 + RT(trS - 1). \label{32}
\end{eqnarray}
Arbitrariness in choosing the signs originating from (\ref{28})
must be fixed in complete analogy with the previous case. Introducing
parametrization
$$K_1 = \omega\cos\phi\cos\theta_1,~~
K_2 = \omega\sin\phi\cos\theta_2, ~~
k_1 = \omega\cos\phi\sin\theta_1, ~~
k_2 = \omega\sin\phi\sin\theta_2,$$
we get from (\ref{29})
\begin{eqnarray}
\omega(\phi,\theta_1,\theta_2 ) &=&
\sqrt{\frac{\left(m_1^2 + m_2^2 \pm  \sqrt{(m_1^2
+ m_2^2)^2 - 4(R^2 + T^2 + RT(trS - 1))}\right)}
{\begin{array}{c}
(s_{23} -s_{32})^2 + \cos^2\phi(s_{13} - s_{31})^2 +
\sin^2\phi(s_{12} - s_{21})^2 \\
+ (s_{31} - s_{13})(s_{12} - s_{21})\sin2\phi \cos(\theta_1 -\theta_2) \\
\end{array} }} \times \nonumber\\
&&\times \sqrt{F_{\pm}\left(R^2 + T^2 + RT(trS - 1)\right)},\label{33}
\end{eqnarray}
\begin{eqnarray}
H &=& \omega(\phi,\theta_1,\theta_2 )
\left(
\begin{array}{c}
- \cos\phi~\cos\theta_1(s_{31}-s_{13}) -
\sin\phi~\cos\theta_2(s_{12}-s_{21})\\
\cos\phi\cos\theta_1(s_{23} - s_{32}) \\
\sin\phi\cos\theta_2(s_{23} - s_{32}) \\
\end{array}
\right),  \label{34} \\
h &=& \omega(\phi,\theta_1,\theta_2 )
\left(
\begin{array}{c}
- \cos\phi~\sin\theta_1(s_{31}-s_{13}) -
\sin\phi~\sin\theta_2(s_{12}-s_{21})\\
\cos\phi\sin\theta_1(s_{23} - s_{32}) \\
\sin\phi\sin\theta_2(s_{23} - s_{32}) \\
\end{array}
\right).  \label{35}
\end{eqnarray}
We have three free parameters that are the angles
$\theta_1,~\theta_2,~\phi.$  The gauge symmetry manifests itself
in the following way. If ${\cal H}(\phi,\theta_1,\theta_2)$ is the
solution of (\ref{24}),
${\cal H}^{\prime} = e^{\imath\alpha}{\cal H}(\phi,\theta_1,\theta_2)
= {\cal H}(\phi,\theta_1 + \alpha,\theta_2 + \alpha)$ is its solution
too.

Unlike the $\lambda = 2$ case, this case contains
extremal configurations
which
are not gauge equivalent to the real ones. Gauge equivalence to the real
field configurations takes place only if $\theta_1 = \theta_2 \equiv
\theta$ :
\begin{eqnarray}
{\cal H}(\phi,\theta,\theta) &=&
\omega(\phi,\theta,\theta)\cos\theta
\left(
\begin{array}{c}
- \cos\phi(s_{31}-s_{13}) -
\sin\phi(s_{12}-s_{21})\\
\cos\phi(s_{23} - s_{32}) \\
\sin\phi(s_{23} - s_{32}) \\
\end{array}
\right),  \nonumber \\
&+&
\imath\omega(\phi,\theta,\theta)\sin\theta
\left(
\begin{array}{c}
- \cos\phi(s_{31}-s_{13}) -
\sin\phi(s_{12}-s_{21})\\
\cos\phi(s_{23} - s_{32}) \\
\sin\phi(s_{23} - s_{32}) \\
\end{array}
\right).  \label{36}
\end{eqnarray}
Let us note that like in the $\lambda = 2$ case, for the solution to
have physical interest it is necessary  that all real parts in
(\ref{36}) have the same signs. This contradicts an analogous
constraint in the $\lambda = 2$ case. Indeed, as it can be seen from
(\ref{25}), if all components of the vector $\epsilon_{ijk}S_{jk}$ have
the same signs, the coordinates of every point on the plane, orthogonal
to it and containing zero point, have different signs. Vice versa,
if the coordinates of any  point belonging to this plane have the
same signs, the coordinates of the vector $\epsilon_{ijk}S_{jk}$ have
different signs.  This situation is illustrated  Fig.1, where the
real solutions of the minimization conditions are depicted. The solution
corresponding to the $\lambda = 2$ case lies on the axis in the Higgs
generation space, around which the matrix $S$ performs rotations.
The solution corresponding to the $\lambda = trS -1$ case lies on the
circle in the plane orthogonal to this axis. The angle $\phi$ in
(\ref{36}) determines the position of the extremum on this circle.
Knowing the expressions for $H$ and $h$ we can obtain from (\ref{24})
the expressions for $\overline{H}$ and $\overline{h}$. Let us note that
after the gauge transformation of (\ref{36}) to the real configuration
the extremal solution for $\overline{\cal H}$ will become real too. It
can be found by using the following formula
\begin{eqnarray} \overline{H} &=&
\sqrt{\frac{\left(m_1^2 + m_2^2 \mp  \sqrt{(m_1^2 + m_2^2)^2 - 4(R^2 +
T^2 + RT(trS - 1))}\right)} {\begin{array}{c}
(s_{23} -s_{32})^2 + \cos^2\phi(s_{13} - s_{31})^2 + \sin^2\phi(s_{12} -
s_{21})^2 \\ + (s_{31} - s_{13})(s_{12} - s_{21})\sin2\phi  \\
\end{array} }} \times \nonumber\\ &&\times \sqrt{F_{\pm}\left(R^2 + T^2
+RT(trS - 1)\right)}\times \nonumber\\
&&\times \frac{R + TS}{\sqrt{R^2 + T^2 + RT(trS - 1)}} \left(
\begin{array}{c} - \cos\phi(s_{31}-s_{13}) -
\sin\phi(s_{12}-s_{21})\\ \cos\phi(s_{23} - s_{32}) \\
\sin\phi(s_{23} - s_{32}) \\ \end{array}
\right).  \nonumber \end{eqnarray}
If (\ref{21}), (\ref{22}) and (\ref{31}), (\ref{32}) are satisfied, both
variants of the solution can occur.  To decide finally which extremum is
suitable for us from the physical point of view, we need to determine
its type. However, yet now we can say that the solution corresponding to
the case $\lambda = trS - 1$ has an additional global symmetry. Indeed,
if ${\cal H}_i$ and ${ \cal \overline{H}}_i$ are the solutions of
(\ref{24}), the field configurations $O_{ij}{\cal H}_j$ and $O_{ij}{\cal
\overline{H}}_j$, where $O$ is some orthogonal matrix commuting with $S$,
will be solutions of (\ref{24}) as well. Breaking this symmetry
generates additional Goldstone bosons, what will be demonstrated
explicitly in the next section.

\section{Higgs masses and  types of extrema}

As it has been noted, we are interested in the extremal field
configurations which are gauge equivalent to the real ones. In this
case, the phases of ${\cal H}$ and ${\cal \overline{H}}$ can be put
equal to zero simultaneously, and we  get
\begin{equation} {\cal H}_i =
v_i + \imath0, ~~~ {\cal \overline{H}}_i = \overline{v}_i + \imath0.
\label{37}
\end{equation}
Let us determine the type of extrema at these
points. To do this, we need to find  the eigenvalues  of
the matrices of second derivatives of the potential
(\ref{7}) at this point. The matrix of second derivatives of (\ref{7})
with respect to the real parts of the neutral $SU(2)$ components $H_i$
and  $\overline{H}_i$ {\Large $$\left( \begin{array}{cc}
\frac{1}{2}\frac{\delta^2 V}{\delta \overline{H}_i \delta
\overline{H}_j}
&
\frac{1}{2}\frac{\delta^2 V}{\delta \overline{H}_i \delta H_j} \\
\frac{1}{2}\frac{\delta^2 V}{\delta H_i \delta \overline{H}_j}
&
\frac{1}{2}\frac{\delta^2 V}{\delta H_i \delta H_j} \\
\end{array} \right)$$}
at the point
$H_i = v_i,~ h_i = 0,~ H^{+}_i = 0,~ \overline{H}_i = \overline{v}_i,~
\overline{h}_i = 0,~ \overline{H}^{-}_i = 0$, has the form \cite{kkk}
\begin{equation}
\left(
\begin{array}{cc}
(m_1^2+x)\delta _{ij}+\frac 12(g^2+g^{\prime}{}^2)\overline v_i\overline v_j &
\mu _{ij}- \frac 12(g^2+g^{\prime}{}^2)\overline v_iv_j \\
\mu_{ji}-\frac 12(g^2+g^{\prime}{}^2)\overline v_j v_i &
(m_2^2-x)\delta _{ij}+\frac 12(g^2+g^{\prime}{}^2)v_iv_j
\end{array} \right).  \label{38}
\end{equation}
The matrix of second derivatives of (\ref{7}) with respect to the
imaginary parts of the neutral $SU(2)$ components $h_i$ and
$\overline{h}_i$ {\Large $$\left(
\begin{array}{cc} \frac{1}{2}\frac{\delta^2 V}{\delta \overline{h}_i
\delta \overline{h}_j}
& \frac{1}{2}\frac{\delta^2 V}{\delta \overline{h}_i \delta h_j} \\
\frac{1}{2}\frac{\delta^2 V}{\delta h_i \delta \overline{h}_j} &
\frac{1}{2}\frac{\delta^2 V}{\delta h_i \delta h_j} \\ \end{array}
\right)$$} at the same point has the form  \cite{kkk}
\begin{equation} \left(
\begin{array}{cc} (m_1^2+x)\delta _{ij} &  -\mu _{ij} \\
-\mu _{ji}            &  (m_2^2-x)\delta _{ij} \end{array}
\right).     \label{39} \end{equation}
And, finally, the matrix of second derivatives of (\ref{7}) with respect
to the charged $SU(2)$ components
$\overline{H}^{-}$ and $\overline{H}^{+}$
{\Large
$$\left( \begin{array}{cc}
\frac{\delta^2 V}{\delta \overline{H}^+_i \delta
\overline{H}^-_j} & \frac{\delta^2 V}{\delta \overline{H}^+_i
\delta H^-_j} \\
\frac{\delta^2 V}{\delta H^+_i \delta
\overline{H}^-_j} & \frac{\delta^2 V}{\delta H^+_i \delta
H^-_j} \\
\end{array} \right)$$} has the form \cite{kkk}:

\begin{equation} \left(
\begin{array}{cc} (m_1^2+z)\delta _{ij}+\frac 12{g^2}\overline
v_i \overline v_j & -\mu _{ij}+ \frac 12 {g^2}\overline v_iv_j \\
-\mu_{ji}+\frac 12{g^2}\overline v_jv_i & (m_2^2-z)\delta _{ij}
+\frac 12{g^2}v_iv_j \end{array} \right), \label{40}
\end{equation}
where $\displaystyle z = \frac{g^{\prime}{}^2 - g^2}{4}(\overline{v}_i^2 -
v_i^2).$
In  complete analogy with the MSSM \cite{haber}, the eigenvalues of
these matrices are the masses of  $CP$-even, $CP$-odd and charged
Higgses \cite{kkk}.  For instance we shall find the eigenvalues of
the mass matrix (\ref{40}) which are the squares of masses of charged
Higgses.  At this moment, we do not fix which variant of the extremum
is realized.  We have only in mind that these vectors $v$ and $\overline{v}$
obey the equations
\begin{eqnarray} -\mu v &=& (m_1^2 + x)\overline{v} \nonumber\\
-\mu^{T} \overline{v} &=& (m_2^2 - x)v.
\label{41} \end{eqnarray}
Now we introduce the matrices
\begin{equation}
u =
\left(
\begin{array}{ccc}
v_1 & 0 & 0 \\
v_2 & 0 & 0 \\
v_3 & 0 & 0 \\
\end{array} \right),
{}~~~ \overline{u} =
\left(
\begin{array}{ccc}
\overline{v}_1 & 0 & 0 \\
\overline{v}_2 & 0 & 0 \\
\overline{v}_3 & 0 & 0 \\
\end{array} \right),
\label{42}
\end{equation}
and write the
following system of linear equations:
\begin{eqnarray}
\left[(m_1^2 + z - \xi)I + \frac{g^2}{2}\overline{u}\overline{u}^{T}\right]f
- \left[\mu - \frac{g^2}{2}\overline{u}u^{T}\right]\overline{f} = 0 \nonumber\\
\left[(m_2^2 - z - \xi)I + \frac{g^2}{2}uu^{T}\right]\overline{f} -
\left[\mu^{T} - \frac{g^2}{2}u\overline{u}^{T}\right]f = 0, \label{43}
\end{eqnarray}
where $f{\rm and}~ \overline{f}$ are some three dimensional vectors.
The system (\ref{43}) has nontrivial solution if corresponding matrix
has determinant equal to zero. To avoid the cumbersome formulae we
denote:
$$m_1 \equiv m_1^2 + x, ~~~m_2 \equiv m_2^2 - x,~~~
A \equiv -\frac{g^2}{2}(\overline{v}^2 - v^2)$$
and absorb the factor $\sqrt{\frac{g^2}{2}}$ in $v$ and $\overline{v}.$
Then, (\ref{43}) take the following form:
\begin{eqnarray}
\left[(m_1 - A - \xi)I +
\overline{u}\overline{u}^{T}\right]f -
\left[\mu - \overline{u}u^{T}\right]\overline{f} = 0 \nonumber\\
\left[(m_2 + A - \xi)I +
uu^{T}\right]\overline{f} -
\left[\mu^{T} - u\overline{u}^{T}\right]f = 0. \label{44}
\end{eqnarray}
It can be shown that if the conditions (\ref{21}), (\ref{22}) and
(\ref{31}), (\ref{32}) are satisfied, we  have
$$det(\mu - \overline{u}u^{T}) \ne 0.$$
Taking this into account we get from (\ref{44})
\begin{equation}
\left[\left((m_2 + A - \xi)I + uu^{T}\right)
\left(\mu - \overline{u}u^{T}\right)^{-1}
\left((m_1 - A - \xi)I + \overline{u}\overline{u}^{T}\right)\right]f
- \left[\mu^{T} - u\overline{u}^{T}\right]f = 0. \label{45}
\end{equation}
The condition for a nontrivial solution for $f$ to exist in (\ref{45}) is
\begin{equation}
det\left[\left((m_2 + A - \xi)I + uu^{T}\right)
\left(\mu - \overline{u}u^{T}\right)^{-1}
\left((m_1 - A - \xi)I + \overline{u}\overline{u}^{T}) -
(\mu^{T} - u\overline{u}^{T}\right)\right] = 0. \label{46}
\end{equation}
Using (\ref{41}), after some transformations we get from (\ref{46})
\begin{eqnarray}
&&det\left[(m_2 + A - \xi)(m_1 - A - \xi)I +
\frac{uu^{T}}{v^2}\left(m_1m_2 - (m_1 - A)(m_2 + A)\right) - \right.
\nonumber\\
&& \left.
- uu^{T}\left(\frac{(2m_1 + v^2)\xi^2 -\xi(m_1 +
m_2)(m_1 + v^2)}{(m_1 + v^2)^2}\right) - \mu\mu^{T}\right] = 0.
\label{47}
\end{eqnarray}
It can be calculated that
\begin{eqnarray}
&&det\left(aI + buu^{T} - RT(S + S^{T})\right) = (a - 2RT)(a - RT(trS -
1))^2 +
\nonumber\\
&& + b~(a^2v^2 -2aRTB_1 + R^2 T^2 (trS - 1) B_2),   \label{48}  \\
&&B_1 \equiv v^2trS - \frac12tr\left(\left(S +
S^{T}\right)uu^{T}\right), \nonumber\\
&&B_2 \equiv v^2(trS + 1) -
tr\left(\left(S + S^{T}\right)uu^{T}\right).  \nonumber
\end{eqnarray}
Here, we have the distinctions between the first and second variants of
the extremum. In the $\lambda = 2$ case, we get $B_1 = B_2 = v^2(trS -
1)$ while in the $\lambda = (trS - 1)$ case $ B_1 = \frac12v^2(trS +
1), ~~~~ B_2 = 2v^2.$ Using (\ref{48}) we can see that a characteristic
equation can be factored in both the cases and following eigenvalues can be
found:  in the $\lambda = 2$ case
\begin{eqnarray}
\xi_1 &=&  \xi_2 =
\frac{1}{2}(m_1^2 + m_2^2) + \Biggl[R^2 +T^2 + RT(trS - 1) \Biggr.
\nonumber\\*
&+& \Biggl.\left(\frac{g^2}{g^2 +
g^{\prime}{}^2}(m_2^2 - m_1^2) \pm \frac{1}{2}\frac{g^2 -
g^{\prime}{}^2}{g^2 + g^{\prime}{}^2}\sqrt{(m_1^2 + m_2^2)^2
-4(R+T)^2}\right)^2 \Biggr]^{\frac{1}{2}}, \nonumber  \\
\xi_3 &=&  \xi_4 =
 \frac{1}{2}(m_1^2 + m_2^2) - \Biggl[R^2 +T^2 + RT(trS - 1)
 \Biggr. \nonumber\\*
&+& \Biggl.\left(\frac{g^2}{g^2 +
g^{\prime}{}^2}(m_2^2 - m_1^2) \pm \frac{1}{2}\frac{g^2 -
g^{\prime}{}^2}{g^2 + g^{\prime}{}^2}\sqrt{(m_1^2 + m_2^2)^2
-4(R+T)^2}\right)^2 \Biggr]^{\frac{1}{2}}, \nonumber \\
\xi_5 &=& m_1^2 + m_2^2 + \frac{g^2}{2}(v^2 +
\overline{v}{}^2) = m_1^2 + m_2^2 + \nonumber \\
&+&
\frac{g^2}{g^2 + g^{\prime}{}^2}(m_1^2 + m_2^2)\frac{\pm (m_1^2 - m_2^2)
-\sqrt{(m_1^2 + m_2^2)^2 -4(R+T)^2}}{\sqrt{(m_1^2 + m_2^2)^2
-4(R+T)^2}}, \nonumber \\ \xi_6 &=& 0 \label{49}
\end{eqnarray} and in the $\lambda = (trS - 1)$ case
\begin{eqnarray} \xi_1 &=&
 \frac{1}{2}(m_1^2 + m_2^2) + \Biggl[(R^2 +T^2 + RT(trS - 1)) \Biggr.
\nonumber\\*
&+& \Biggl.\left(\frac{g^2}{g^2 + g^{\prime}{}^2}(m_2^2 -
m_1^2) \pm \frac{1}{2}\frac{g^2 - g^{\prime}{}^2}{g^2 +
g^{\prime}{}^2}\sqrt{(m_1^2 + m_2^2)^2 -4(R^2 +T^2 + RT(trS -
1))}\right)^2 \Biggr]^{\frac{1}{2}}, \nonumber \\
\xi_2 &=&
 \frac{1}{2}(m_1^2 + m_2^2) - \Biggl[(R^2 +T^2 + RT(trS - 1)) \Biggr.
\nonumber\\*
&+& \Biggl.\left(\frac{g^2}{g^2 + g^{\prime}{}^2}(m_2^2 -
m_1^2) \pm \frac{1}{2}\frac{g^2 - g^{\prime}{}^2}{g^2 +
g^{\prime}{}^2}\sqrt{(m_1^2 + m_2^2)^2 -4(R^2 +T^2 + RT(trS -
1))}\right)^2 \Biggr]^{\frac{1}{2}}, \nonumber \\
\xi_3 &=&
 \frac{1}{2}(m_1^2 + m_2^2) + \Biggl[(R +T)^2 \Biggr. \nonumber\\*
&+& \Biggl.\left(\frac{g^2}{g^2 + g^{\prime}{}^2}(m_2^2 - m_1^2) \pm
\frac{1}{2}\frac{g^2 - g^{\prime}{}^2}{g^2 + g^{\prime}{}^2}\sqrt{(m_1^2 +
m_2^2)^2 -4(R^2 +T^2 + RT(trS - 1))}\right)^2 \Biggr]^{\frac{1}{2}},
\nonumber  \\
\xi_4 &=&
 \frac{1}{2}(m_1^2 + m_2^2) - \Biggl[(R +T)^2 \Biggr. \nonumber\\*
&+& \Biggl.\left(\frac{g^2}{g^2 + g^{\prime}{}^2}(m_2^2 - m_1^2) \pm
\frac{1}{2}\frac{g^2 - g^{\prime}{}^2}{g^2 + g^{\prime}{}^2}\sqrt{(m_1^2 +
m_2^2)^2 -4(R^2 +T^2 + RT(trS - 1))}\right)^2 \Biggr]^{\frac{1}{2}},
\nonumber \\
\xi_5 &=& m_1^2 + m_2^2 + \frac{g^2}{2}(v^2 +
\overline{v}{}^2) = m_1^2 + m_2^2 + \nonumber \\
&+&
\frac{g^2}{g^2 + g^{\prime}{}^2}(m_1^2 + m_2^2)\frac{\pm (m_1^2 - m_2^2)
-\sqrt{(m_1^2 + m_2^2)^2 -4(R^2 +T^2 + RT(trS - 1))}}{\sqrt{(m_1^2 + m_2^2)^2 -
4(R^2 +T^2 + RT(trS - 1))}}, \nonumber\\
\xi_6 &=& 0 .\label{50}
\end{eqnarray}
Acting in the same manner we can find the eigenvalues for the mass
matrix of $CP$-even Higgses (\ref{38}) in the $\lambda = 2$ case
\begin{eqnarray}
\xi_1 &=&
\frac{\pm(m_1^2 - m_2^2)(m_1^2 + m_2^2)}{2 \sqrt{(m_1^2 +m_2^2)^2 -
4(R+T)^2}}
+ \Biggl[\frac{(m_1^2 - m_2^2)^2 (m_1^2 + m_2^2)^2}{4((m_1^2 +
m_2^2)^2 -4(R+T)^2)} \Biggr. \nonumber\\*
&+& \Biggl. (m_1^2 + m_2^2)^2 - 4(R+T)^2
\mp (m_1^2 - m_2^2)\sqrt{(m_1^2 + m_2^2)^2 -
4(R+T)^2}\Biggr]^{\frac{1}{2}}, \nonumber \\
\xi_2 &=&
\frac{\pm(m_1^2 - m_2^2)(m_1^2 + m_2^2)}{2 \sqrt{(m_1^2 +m_2^2)^2 -
4(R+T)^2}}
- \Biggl[\frac{(m_1^2 - m_2^2)^2 (m_1^2 + m_2^2)^2}{4((m_1^2 +
m_2^2)^2 -4(R+T)^2)} \Biggr. \nonumber\\*
&+& \Biggl.(m_1^2 + m_2^2)^2 - 4(R+T)^2
\mp (m_1^2 - m_2^2)\sqrt{(m_1^2 + m_2^2)^2 -
4(R+T)^2}\Biggr]^{\frac{1}{2}}, \nonumber \\
\xi_3 &=& \xi_4 = \frac{1}{2}\left((m_1^2 + m_2^2) + \sqrt{(m_1^2
+ m_2^2)^2 - 4RT(3 - trS)}\right), \nonumber\\
\xi_5 &=& \xi_6 = \frac{1}{2}\left((m_1^2 + m_2^2) - \sqrt{(m_1^2
+ m_2^2)^2 - 4RT(3 - trS)}\right)  \label{51}
\end{eqnarray}
and in the $\lambda = trS - 1$ case:
\begin{eqnarray}
\xi_1 &=&
\frac{\pm(m_1^2 - m_2^2)(m_1^2 + m_2^2)}{2 \sqrt{(m_1^2 +m_2^2)^2 -
4(R^2 + T^2 + RT(trS - 1))}}  \nonumber\\
&+& \Biggl[\frac{(m_1^2 - m_2^2)^2 (m_1^2 + m_2^2)^2}{4((m_1^2 +
m_2^2)^2 -4(R^2 + T^2 + RT(trS - 1)))} \Biggr. \nonumber\\*
&+& (m_1^2 + m_2^2)^2 - 4(R^2 + T^2 + RT(trS - 1)) \nonumber\\*
&\mp& \Biggl. (m_1^2 -m_2^2)\sqrt{(m_1^2 + m_2^2)^2 -
4(R^2 + T^2 + RT(trS - 1))} \Biggr]^{\frac{1}{2}},
\nonumber \\
\xi_2 &=&
\frac{\pm(m_1^2 - m_2^2)(m_1^2 + m_2^2)}{2 \sqrt{(m_1^2 +m_2^2)^2 -
4(R^2 + T^2 + RT(trS - 1))}} \nonumber\\
&-& \Biggl[\frac{(m_1^2 - m_2^2)^2 (m_1^2 + m_2^2)^2}{4((m_1^2 +
m_2^2)^2 -4(R^2 + T^2 + RT(trS - 1)))} \Biggr. \nonumber\\*
&+& (m_1^2 + m_2^2)^2 - 4(R^2 + T^2 + RT(trS - 1)) \nonumber\\*
 &\mp& \Biggl. (m_1^2 -m_2^2)\sqrt{(m_1^2 + m_2^2)^2 -
4(R^2 + T^2 + RT(trS - 1))} \Biggr]^{\frac{1}{2}},
\nonumber \\
\xi_3 &=& \frac{1}{2}\left((m_1^2 + m_2^2) + \sqrt{(m_1^2
+ m_2^2)^2 + 4RT(3 - trS)}\right), \nonumber\\
\xi_4 &=& \frac{1}{2}\left((m_1^2 + m_2^2) - \sqrt{(m_1^2
+ m_2^2)^2 + 4RT(3 - trS)}\right), \nonumber\\
\xi_5 &=& m_1^2 + m_2^2, \nonumber\\
\xi_6 &=& 0,     \label{52}
\end{eqnarray}
and eigenvalues for the mass matrix of $CP$-odd Higgses (\ref{39})
in the $\lambda = 2$ case:
\begin{eqnarray}
\xi_1 &=& 0, \nonumber\\
\xi_2 &=& m_1^2 + m_2^2, \nonumber\\
\xi_3 &=& \xi_4 = \frac{1}{2}\left((m_1^2 + m_2^2) + \sqrt{(m_1^2
+ m_2^2)^2 - 4RT(3 - trS)}\right),\nonumber\\
\xi_5 &=& \xi_6 = \frac{1}{2}\left((m_1^2 + m_2^2) - \sqrt{(m_1^2
+ m_2^2)^2 - 4RT(3 - trS)}\right) \label{53}
\end{eqnarray}
and in the  $\lambda = trS - 1$ case
\begin{eqnarray}
\xi_1 &=& \xi_2 =0, \nonumber\\
\xi_3 &=& \xi_4 = m_1^2 + m_2^2, \nonumber\\
\xi_5 &=&  \frac{1}{2}\left((m_1^2 + m_2^2) + \sqrt{(m_1^2
+ m_2^2)^2 + 4RT(3 - trS)}\right), \nonumber\\
\xi_6 &=& \frac{1}{2}\left((m_1^2 + m_2^2) - \sqrt{(m_1^2
+ m_2^2)^2 + 4RT(3 - trS)}\right). \label{54}
\end{eqnarray}

Let us consider the expressions for the eigenvalues of
$CP$-even Higgses mass matrix (\ref{38}). It is easy to show that if
the conditions (\ref{21}), (\ref{22}) and (\ref{31}), (\ref{31}) are
satisfied, type of the extremum  depends on  $sign(RT).$  In fact, if
$RT > 0$, the matrix (\ref{38})  has all non-negative eigenvalues
(\ref{51}) in the $\lambda = 2$ case while another extremum is the
saddle point. Otherwise, if $RT < 0$, the matrix (\ref{38})  has
all non-negative eigenvalues (\ref{52}) in the $\lambda = trS - 1$
case while extremum corresponding to the $\lambda = 2$ case proves to be
saddle point. We can observe the same situation considering eigenvalues
of the matrices (\ref{39}) and (\ref{40}). Having written the matrix of
second derivatives of the potential (\ref{7}) at zero, we can see that
zero is the saddle point for any sign of $RT$ (if the conditions (\ref{21}),
(\ref{22}) and (\ref{31}), (\ref{32}) are satisfied).  Goldstone bosons,
which we have in the $\lambda = 2$ case, are the results of electroweak
symmetry breaking. They generate masses of gauge $Z$-boson ($CP$-odd
Goldstone boson) and $W^{\pm}$ bosons (charged Goldstone bosons). The
additional zero eigenvalues in the $\lambda = trS - 1$ case correspond to the
global symmetry breaking. As we have written in the previous section, this
global symmetry is the symmetry of the potential (\ref{7}) with respect to
the rotation of fields in the Higgs generation space which is
performed by the
orthogonal matrices commuting with $S$. Taking into account the presence
of these additional Goldstone bosons, we may conclude that this extremum
is not suitable from the physical point of view. For final conclusion to
make, let us calculate the significance of the potential (\ref{7}) on the
extremal configurations. In both the cases we get
\begin{equation} V_{ext} = -\frac{2}{g^2 + g^{\prime}{}^2}x^2.
\label{55}
\end{equation}
Then, using (\ref{17}) and (\ref{28}), we calculate
\begin{equation}
V_{ext}^{\lambda = 2} = -\frac{1}{2(g^2 + g^{\prime}{}^2)}\left(|m_1^2 -
m_2^2| - \sqrt{(m_1^2 + m_2^2)^2 - 4(R + T)^2}\right)^2, \label{56}
\end{equation}
\begin{equation}
V_{ext}^{\lambda = trS - 1} = -\frac{1}{2(g^2 +
g^{\prime}{}^2)}\left(|m_1^2 - m_2^2| - \sqrt{(m_1^2 + m_2^2)^2 - 4(R^2 +
T^2 + RT(trS - 1))}\right)^2. \label{57}
\end{equation}
Thus, for $RT > 0$ the absolute minimum of the potential (\ref{7})
is the extremum corresponding to $\lambda = 2$, while zero and extremum
corresponding to  $\lambda = trS - 1$ are the saddle points. In the opposite
case, for $RT < 0$ the absolute minimum is the extremum corresponding to
$\lambda = trS - 1$ while zero and extremum corresponding to the $\lambda = 2$
case are the saddle points. We must discard the second extremum due to
additional Goldstone bosons. Taking into account the afore-mentioned
arguments and conclusions of the previous section, we summarize that
the potential (\ref{7}) has an absolute minimum with respect to neutral
components of the $SU(2)$ scalar Higgs doublets, interesting physically,
on the field configurations (\ref{20}) and (\ref{23}) under the
following restrictions on the quantities of the potential (\ref{7}):
\begin{eqnarray}
&& R + T < 0, \nonumber\\
&& \epsilon_{ijk}S_{jk} > 0 ~{\rm for~~any}~~i,  \nonumber\\
&& m_1^2 + m_2^2 > 2|R + T|, \nonumber\\
&& m_1^2m_2^2 < (R + T)^2, \nonumber\\
&& RT > 0. \nonumber
\end{eqnarray}
Finally, let us make some remarks regarding extremal field configurations
in (\ref{34}), (\ref{35}) discarded by us in view of their
inequivalency to the real ones. The potential (\ref{7}) on these configurations
equals the significance (\ref{57}), which  is greater than the absolute minimum
(\ref{56}). Moreover, if our system is in the vicinity of this extremum,
the afore-mentioned global symmetry in the generation space is broken and
additional Goldstone bosons appear.

\section{Summary}
The main reason why we have succeeded in the exact solution of
the system (\ref{8}) is that this system includes nonlinearity as a whole
having the form of a quadratic combination of unknowns. This type of
nonlinearity is generated by quartic terms in the potential (\ref{7}),
which, in their turn, arise after excluding auxiliary
non-dynamical components of gauge supermultiplets. Therefore, this form
of quartic terms in the potential is typical of the $N =1$
supergravity GUT's with enlarged  Higgs sector \cite{ib}.
The quadratic part of the potential (\ref{7}) has a specific form dictated
by finiteness. This fact allowed us to find the nonlinear combination in
(\ref{8}) explicitly. Let us note also that, as it is not difficult to
see, our result for absolute minimum configurations (\ref{20}) and
(\ref{23}) is a simple generalization of the analogous result in the
MSSM \cite{haber}.

In conclusion, we would like to attract  attention to the
interesting phenomenological predictions for the quark mass spectrum.
After the transition of the system to the absolute minimum (\ref{56}) of the
potential on the field configurations (\ref{20}) and (\ref{23}), we fix
the phases of these configurations to equal zero, and the following relations
between up and down quark masses  can be observed:
$$\frac{m_u}{m_d} = \frac{m_c}{m_s} = \frac{m_t}{m_b}.$$
Quark masses in these relations are running masses and must be taken on
the $M_Z$ scale [\cite{kaz}, \cite{di}]. Approximate estimations show
that this type of a relation between up and down quark masses can take place
\cite{di}. At the same time, the hierarchy between quark generations
is completely controlled by the matrix $S$ that is the parameter of the
theory. Parametrizing this orthogonal matrix  by three Eiler angles
$\theta_1, \theta_2, \theta_3$, we can get the following hierarchy
relations between up quarks:
$$m_u : m_c : m_t = \cos\frac{\theta_1}{2}~\sin\frac{\theta_2 +
\theta_3}{2} :  \sin\frac{\theta_1}{2}~\sin\frac{\theta_2 -
\theta_3}{2} :  \sin\frac{\theta_1}{2}~\cos\frac{\theta_2 -
\theta_3}{2}.$$
It is clear that we can fit these angles in order to guarantee any
hierarchy. Unfortunately, we have not succeeded connecting $S$
with other parameters of the theory. A complete analysis of this model with
numerical results for masses of all particles of the theory is in
preparation \cite{kkk} and will be published elsewhere.
\vspace{1cm} \\
{\bf Acknowledgments} \\
I want to thank sincerely D.I.~Kazakov and M.Yu.~Kalmykov for their
close collaboration and useful conversations. I am indebted to
L.V.~Avdeev for helpful discussions.

\begin{center} 
\end{center}
\pagebreak
Figure 1:~The real solutions of the minimization
conditions. The solution corresponding to the $\lambda = 2$
case lies on the axis in the Higgs generation space, around which the
matrix $S$ performs rotations.  The solution corresponding to the
$\lambda = trS -1$ case lies on the circle in the plane orthogonal to
this axis. The angle $\phi$ in (\ref{36}) determines the position of the
extremum on this circle.
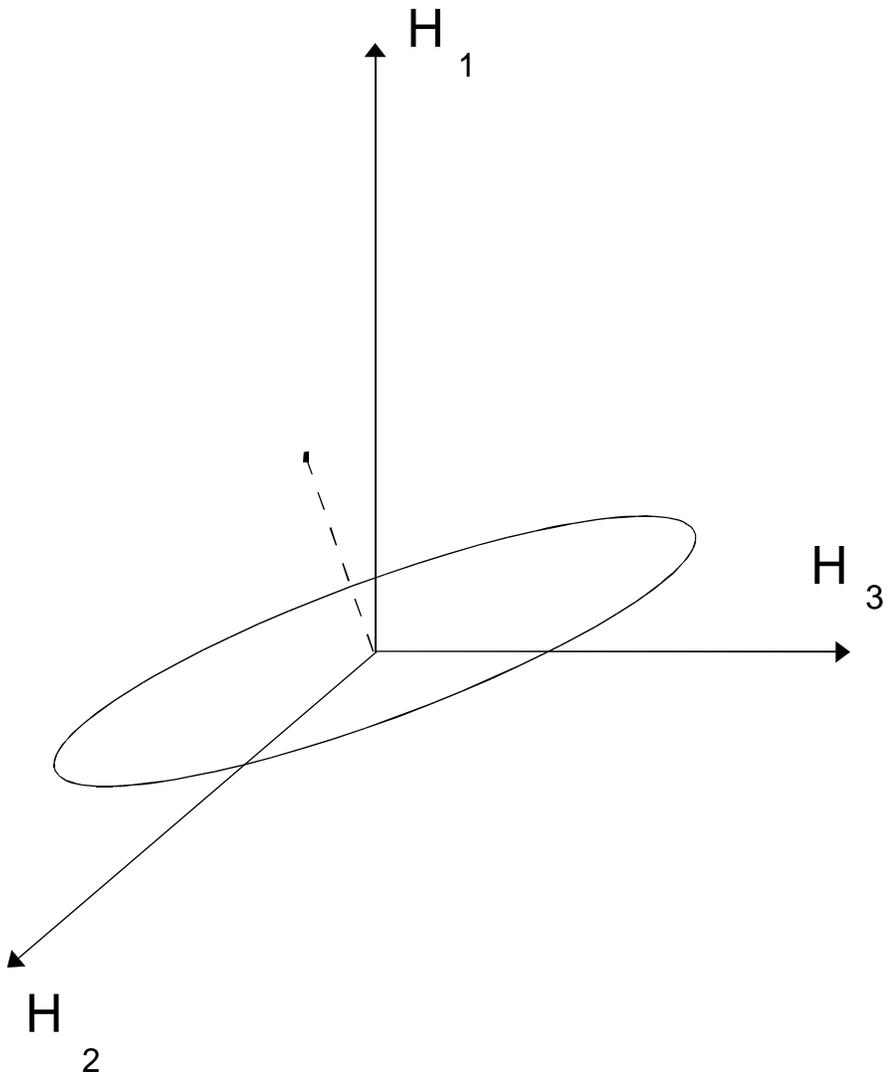
\end{document}